\documentclass[11pt]{article}
\usepackage[english]{babel}
\usepackage[utf8]{inputenc}
\usepackage[nottoc]{tocbibind}
\usepackage{graphicx}
\usepackage{amsmath}
\usepackage{amssymb}
\usepackage{amsxtra}

\title{Numerical simulation of dark atom interaction with nuclei}
\author{T.E. Bikbaev$^{1}$, M.Yu. Khlopov$^{1,2,3}$, A.G. Mayorov$^{1}$\\
$^{1}$ National Research Nuclear University MEPhI \\(Moscow Engineering Physics Institute),\\ 115409 Moscow, Russia\\
$^{2}$ Institute of Physics, Southern Federal University\\ Stachki 194 Rostov on Don 344090, Russia\\
$^{3}$  Université de Paris, CNRS, Astroparticule et Cosmologie,\\ F-75013 Paris, France,\\ e-mail khlopov@apc.univ-paris.fr}

\begin{document}
\maketitle

\begin{abstract}
The old and still not solved problem of dark atom solution for the puzzles of direct dark matter searches is related with rigorous prove of the existence of a low energy bound state in the dark atom interaction with nuclei. Such prove must involve a self-consistent account of the nuclear attraction and Coulomb repulsion in such interaction. In the lack of usual small parameters of atomic physics like smallness of electromagnetic coupling of the electronic shell or smallness of the size of nucleus as compared with the radius of the Bohr orbit the rigorous study of this problem inevitably implies numerical simulation of dark atom interaction with nuclei. Our approach to such simulations of $O$He~--nucleus interaction involves multi-step approximation to the realistic picture by continuous addition to the initially classical picture of three point-like body problem essential quantum  mechanical features.
\end{abstract}



\section{Introduction}\label{s:intro}
According to the modern cosmology, dark matter is non-baryonic and is associated with physics that has not yet been sufficiently studied and, in fact, unknown to us. If it consists of particles, then they are predicted beyond the Standard Model. To be considered as candidates for dark matter these particles  must satisfy a set of conditions: they must be stable, must explain the measured dark matter density, and decouple from plasma and radiation, at least before the  beginning of the matter dominated stage \cite{ansoldi2006proceedings,bennett2007proceedings}. The easiest way to satisfy the above conditions is to assume the existence of neutral, elementary Weakly Interacting Massive Particles (WIMP). However, the results of the WIMP searches are contradictory and the existing uncertainty in the choice of “dark” particles has given rise to many different models suggesting various objects for the role of dark matter candidates \cite{Bertone_2005,KHLOPOV_2013,Khlopov_2017,scott2011searches}. In these models, new particles should possess some new fundamental symmetry and the corresponding conserved charge in order to protect their stability \cite{Khlopov_2017,fabbrichesi2020dark,Khlopov_2019}.

An important problem for scenarios of hypothetical, stable, electrically charged particles is their absence in the matter around us. If they exist, they should be present in the ordinary matter in the form of anomalous isotopes (with an anomalous Z/A ratio). The main difficulty for these scenarios is the suppression of the abundance of positively charged particles bound with electrons, which behave like anomalous isotopes of hydrogen or helium. Serious experimental restrictions on such isotopes, especially on anomalous hydrogen, very severely limit the possibility of stable positively charged particles \cite{Cudell:2012fw}. 

This problem is also unsolvable if the model predicts stable particles with charge $-1$. Such particles bind with primordial helium in $+1$ charged ions, which recombine with electrons in atoms of anomalous hydrogen \cite{Khlopov:2015nrq}. 

In this connection, stable negatively charged particles can only have charge $-2$ ~-- we will denote them by $O^{--}$ or in the general case even charge $-2n$, where $n$ is any natural number. 

In the present paper, we consider a scenario of  composite dark matter, in which hypothetical stable $O^{--}$ particles avoid experimental discovery, because they form neutral atom-like states $O$He with primordial helium, called “dark” atoms \cite{khlopov2019conspiracy}. Since all these models also predict the corresponding $+2$ charged antiparticles, the cosmological scenario should provide a mechanism for their suppression, which, naturally, can take place in the charge-asymmetric case corresponding to an excess of $ -2 $ charged particles $O^{--}$ \cite{KHLOPOV_2013} \footnote{Electric charge of this excess is compensated by the corresponding excess of positively charged baryons so that the electroneutrality of the Universe is preserved}. Then their positively charged antiparticles can effectively annihilate in the early universe. There are various models in which such stable $-2$ charged particles are predicted \cite{belotsky2006composite,Khlopov_2006,Khlopov_2008}.

\section{"Dark" atoms $O$He}
“Dark” atom is the bound system of $O^{--}$ particle and $^4$He nucleus. In the approximation of our current numerical model, $\alpha$-particle is point-like and moves along the Bohr radius. Then the binding energy of $O$He for a point charge of $^4$He is given by:
\begin{equation}
I_{0}=\cfrac{Z_{O^{--}}^2Z_{He}^2\alpha^2m_{He} }{2}\approx1.6\hspace{1.5mm}\text{MeV,}   
\label{eq}  
\end{equation}
where $\alpha$ --~ is a fine structure constant, $Z_{O^{--}}$ and $Z_{He}$ --~ electric charges of $O^{--}$ particle and nuclei He respectively, $m_{He}$ --~ is the $\alpha$-particle mass.

The Bohr radius of $He$ rotation in “dark” $O$He atoms is equal to \cite{Khlopov:2010ik}:
\begin{equation}
R_{b}=\cfrac{\hbar c}{Z_{O^{--}} Z_{He} m_{He} \alpha}\approx2\cdot10^{-13}\hspace{1.5mm}\text{cm} 
\label{eq}  
\end{equation}

In all models of $O$-helium, $O^{--}$ behaves like a lepton or as a specific cluster of heavy quarks of new families with suppressed hadron interaction \cite{khlopov2005composite}. Therefore, the strong interaction of $O$He with matter is determined by the nuclear interaction of $He$. The mass $O^{--} $, $ m_{O^{--}}$, is the only free parameter of new physics. The experimental search at the LHC for stable doubly charged particles gives a lower limit for their mass about $1 \text{TeV}$ \cite{beylin2020new}. 

The neutral primordial nuclear-interacting objects, that is, “dark” $O$He atoms, dominate in the modern density of nonrelativistic matter and play the role of a non-trivial form of strongly interacting dark matter. The active influence of this type of dark matter on nuclear transformations requires special research and development of the nuclear physics of $O$-helium. This is especially important for a quantitative assessment of the role of “dark” atoms in primordial cosmological nucleosynthesis and in the evolution of stars \cite{Khlopov:2010ik}.

The importance of the O-helium hypothesis is that it can explain the conflicting results of a direct search for dark matter, due to the specifics of the interaction of “dark” atoms with the matter of underground detectors \cite{BERNABEI_2013}. Namely, positive results on the detection of dark matter particles in experiments such as DAMA / NaI and DAMA / LIBRA, which seem to contradict all other experiments, for example, with XENON100, LUX, CDMS, which give a negative result.

One of the main problems with the "dark" $O$He atoms is that their constituents can interact too strongly with ordinary matter. This is because $O$-helium, although neutral, initially has an unshielded nuclear attraction to the nuclei of matter. Which can lead to the destruction of the bound $O$He system and the formation of anomalous isotopes. In turn, there are very strong experimental limitations on the concentration of these isotopes in the terrestrial soil and sea water \cite{Cudell:2012fw}. To avoid this problem, it is assumed that the effective potential of $O$He-nucleus interaction will have a barrier preventing the merging of He and/or $O^{--}$ with nucleus. Under these conditions, “dark” atoms interaction with matter doesn't lead to anomalous isotopes overproduction, which is the key point for the $O$-helium hypothesis.

In this work, a description of the performed numerical simulation of the interaction of “dark” $O$-helium atoms with the nuclei of baryonic matter is given with the aim to explore the conditions for the existence of their low-energy bound state, which can explain positive results of DAMA/NaI and DAMA/LIBRA experiments. Within the framework of the proposed approach to such modeling, in order to reveal the essence of the processes of nuclear interaction of $O$He with nuclei of baryonic matter, the approach is based on the classical model, where the effects of quantum physics are gradually introduced.

\section{Numerical modeling of the interaction of $O$He with the nucleus of baryonic matter.}
\subsection{Modeling $O$He.}
To model the "dark" atom of $O$-helium (the $O$He system) was considered, consisting of two point-like, bound particles: the He nucleus and the $O^{--}$ particle. A spherical coordinate system was introduced, at the center of which the particle $O^{--}$ is meant, and around it along the surface of the sphere, the radius of which is equal to the radius of the atom $O$He $R_{b}$ (see formula (2)) the He nucleus moves stochastically, with a constant Bohr velocity $V_{\alpha}$.
The speed $V_{\alpha}$ is:
\begin{equation}
V_{\alpha}=\cfrac{\hbar c^{2}}{m_{He} R_{b}}\approx3\cdot10^{4} \cfrac{\text{cm}}{\text{s}}
\label{eq}  
\end{equation}

The initial task in modeling the interaction of $O$He with nuclei was to construct a numerical model of $O$-helium, which would allow to describe the motion of an $\alpha$-particle around  $O^{--}$. It should be used in the main numerical model, in which the motion and interaction of $O$He with the nuclei will be simulated.

Let us consider how the $O$He system was modeled (see Figure 1).
\begin{figure}[h!]
\centering
\includegraphics[scale=0.55]{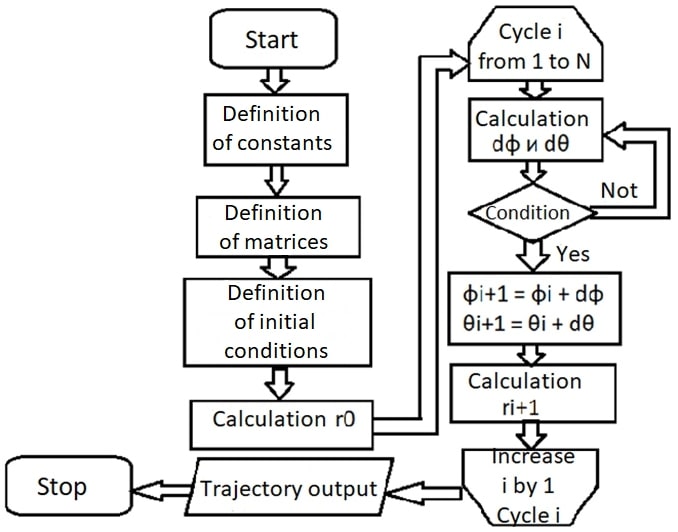}
\caption{Block diagram of the $O$He system simulation}
\label{fig:Ris18}
\end{figure}

An $\alpha$-particle in the bound $O$He system has only two independent degrees of freedom, which are taken as the polar and azimuthal angles. Its Cartesian coordinates are expressed through the projections $R_{b}$, that is, the components of the radius vector at each moment of time, in order to use them to construct the trajectory of the $\alpha$-particle. In Figure 1, the defined matrices mean all the quantities necessary to describe the motion of the $\alpha$-particle, that is, its polar $\theta$ and azimuthal $\phi$ angles, as well as the change in these angles ($d\theta$ and $d\phi$) and components of the radius vector $r$.

$\phi_{0}$ and $\theta_{0}$ in Figure 1 are the initial values of the angles through which the initial components of the radius vector of the $\alpha$-particle $r_{0}$ are calculated.

Changes in the polar d$\theta$ and azimuthal d$\phi$ angles are defined as follows:
\begin{equation}
     d\theta=\biggl({\cfrac{V_{\alpha}dt}{R_{b}}}\biggr)\biggl(2rand-1\biggr)
    \label{eq}  
\end{equation}
\begin{equation}
    d\phi=\cfrac{\sqrt{\biggl({\cfrac{V_{\alpha}dt}{R_{b}}}\biggr)^2-\biggl(d\theta\biggr)^2}}{\cos\bigl({\theta}\bigr)}\biggl(2rand-1\biggr)
    \label{eq}
\end{equation}
where $ rand $ is a random variable with a uniform distribution over the range from 0 to 1.

The condition in Figure 1 means the following inequality:
\begin{equation}
     \biggl(d\theta\biggr)^2+\biggl(\cos\theta d\phi\biggr)^2\leq \biggl({\cfrac{V_{\alpha}dt}{R_{b}}}\biggr)^2
    \label{eq}  
\end{equation}
The physical meaning of this condition is that the square of the distance traveled by an $\alpha$-particle in time $dt$ over the surface of a sphere of radius $R_{b}$ with a constant velocity $V_{\alpha}$ cannot be less than the sum of the squares of the distances covered for that the same time over the surface of a sphere of the same radius with the same velocity in the polar and azimuthal directions.

In general, from Figure 1 it is clear that in each iteration, changes in the azimuthal and polar angles are determined, which are added to their old values ($\phi_{i} $ and $\theta_{i}$) and using the new angles obtained ($\phi_{i+1}$ and $\theta_{i+1}$) the following components of the radius vector of the $\alpha$-particle $r_{i+1}$ are calculated.

As a result, according to the obtained data, written in the matrix containing the values of the components of the radius vector of the $\alpha$-particle at each moment of time $r$, the program builds its trajectory along the surface of a sphere of the Bohr radius $R_{b}$ (Figure 2). Figure 2 shows a sphere of radius $R_{b}$, on the surface of which the red dots mark the location of the $\alpha$-particle between times $dt$. Filling the sphere with dots depends on the number of loop iterations, that is, if there are too many of them, the sphere will be densely filled with dots.
\begin{figure}[h!]
\centering
\includegraphics[scale=0.4]{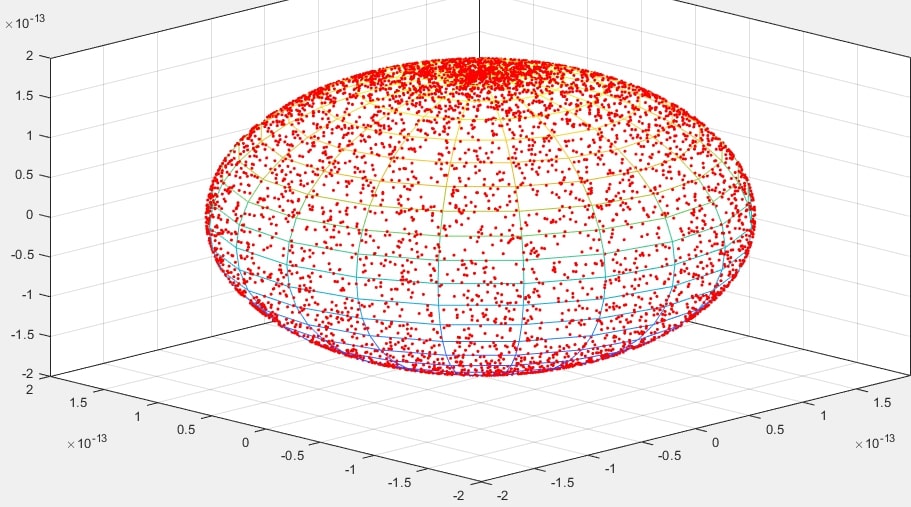}
\caption{The density of the distribution of the coordinates of the $\alpha$-particle on the surface of the sphere of the Bohr radius $R_{b}$}
\label{fig:Ris18}
\end{figure}
\subsection{The coordinate system of the $OHe$–nucleus system.}
Before we start modeling the $OHe$ system and the nucleus of baryonic matter, taking into account all the forces acting between particles, that is, modeling the interaction of three bodies, let us consider the coordinate system for the $OHe$--nucleus system.

The system $OHe$-nucleus consists of three charged, pointlike (in this work) particles, in which a linked system of two other particles moves to one particle “fixed” at the center of coordinates. The particle at the origin is target nucleus of the baryonic matter, and the moving particles mean the $\alpha$-particle and the $O^{--}$. In this case, the $\alpha$-particle rotates along the Bohr radius $R_{b}$ around the particle $O^{--}$.

In order to describe the trajectories of motion of the $\alpha$-particle and the $O^{--}$ consider a spherical coordinate system with a point target nucleus $A$ at the origin of the coordinate system. It introduces the radius vector (see Figure 3) of the $O^{--}$ $\vec{r}$ and the radius vector of the $\alpha$-particle $\vec{r}_{\alpha}$.
Wherein:
\begin{equation}
    \vec{r}_{\alpha}=\vec{r}+\vec{R}_{b}
    \label{eq}
\end{equation}
Accordingly, for the radius vector of the $\alpha$-particle and the $O^{--}$ azimuthal ($\phi_{\alpha}$ and $\phi_{O^{--}}$) and polar ($\theta_{\alpha}$ and $\theta_{O^{--}}$) angles. Figure 3 also shows the particle velocity vector $O^{--}$, $\vec{V}$, the angle between $\vec{V}$ and the horizontal line, $\alpha$, and the initial coordinates of the particle $O^{--}$ $[X_{0}, Y_{0}, Z_{0}]$.
\begin{figure}[h!]
\centering
\includegraphics[scale=0.30]{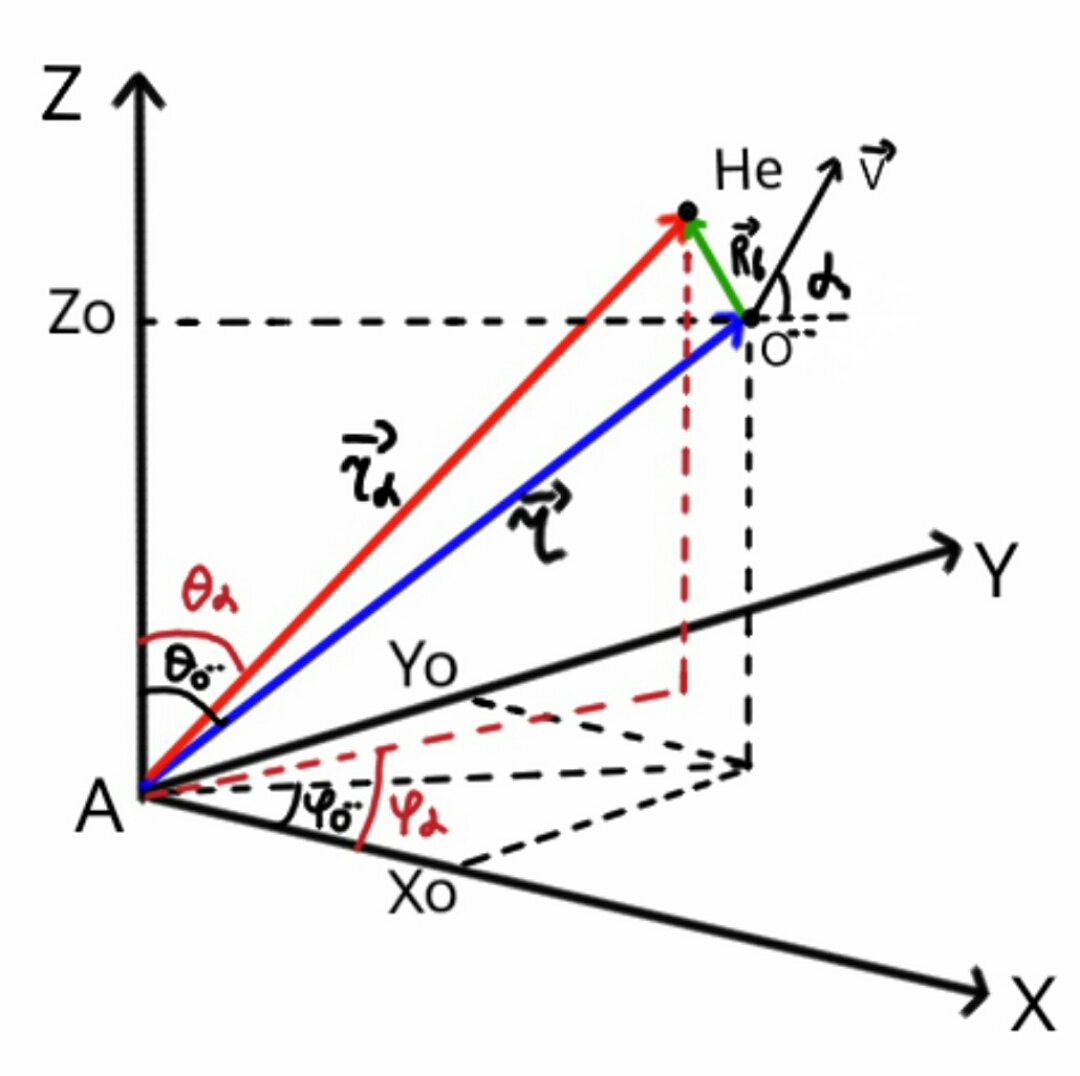}
\caption{Coordinate system $OHe$--core.}
\label{fig:lol3}
\end{figure}
Before proceeding to the description of modeling the $OHe$ system and the nucleus of baryonic matter, taking into account interactions between particles, it should be said that it is possible to construct the effective potential between $O$-helium and the nucleus of baryonic matter (see Figure 4). This potential includes electromagnetic and nuclear interactions. And the task of modeling is precisely to introduce these interactions in order to reproduce the effects of this potential numerically.
\begin{figure}[h!]
\centering
\includegraphics[scale=0.7]{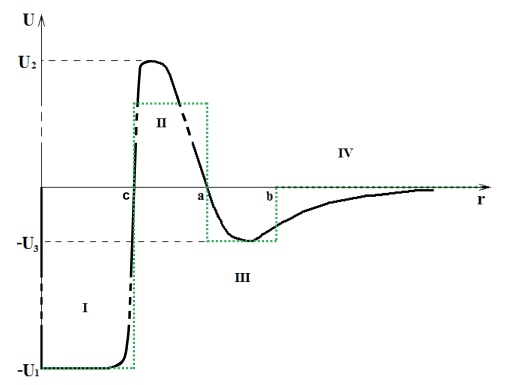}
\caption{Effective potential between $OHe$ and the nucleus of baryonic matter \cite{Khlopov:2010ik}}
\label{fig:Ris17}
\end{figure}
\subsection{Coulomb interaction in the $OHe$~--nucleus system}
At this stage of modeling, a system of three point systems interacting with each other through the Coulomb forces of charged particles is considered, with the above choice of the coordinate system.

A Coulomb interaction acts between the $\alpha$-particle and the target nucleus in the considered coordinate system, which is determined by the force:
\begin{equation}
    \Vec{F}^{e}_{Z\alpha}=\Vec{F}^{e}_{Z\alpha}(\vec{r}_{\alpha})=\cfrac{ZZ_{\alpha}e^2\vec r_{\alpha}}{r_{\alpha}^3},
    \label{eq}
\end{equation}
where $Z$ is the charge of nucleus. Coulomb interaction between the $O^{--}$ particle and the target nucleus, which is determined by the force:
\begin{equation}
    \Vec{F}^{e}_{ZO}=\Vec{F}^{e}_{ZO}(\vec{r})=\cfrac{ZZ_{O}e^2\vec r}{r^3}.
    \label{eq}
\end{equation}
The task of this stage was to simulate the interaction, by means of Coulomb forces (8) and (9), in the coordinate system $OHe$~--nucleus, where the motion of the $He$ nucleus in the bound state $ OHe $ is described according to the algorithm presented in the previous section.

The simulation was carried out as follows (see Figure 5).
\begin{figure}[h!]
\centering
\includegraphics[scale=0.7]{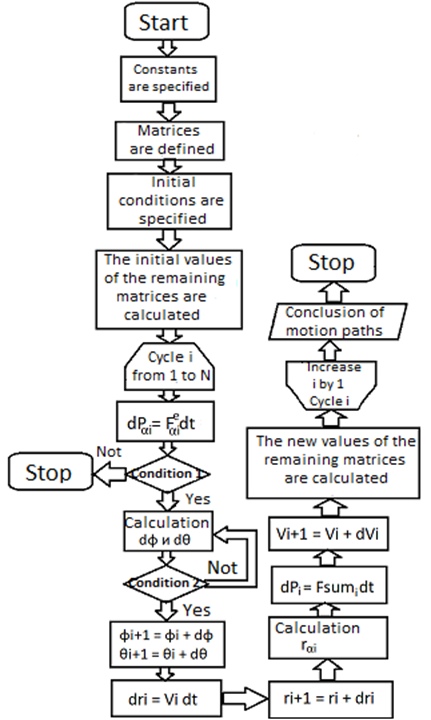}
\caption{Block diagram for modeling the Coulomb interaction in the $OHe$~--nucleus system}
\label{fig:lol5}
\end{figure}

We use the following initial conditions: the initial coordinates of $O^{--}$ $[x_{0}, y_{0}, z_{0}]$ (or $r_{0}$) and the initial components of its velocity $[V_{x_{0}}, V_{y_{0}}, V_{z_{0}}]$ (or $V_{0}$). Then the initial values of all previously determined values are calculated.

Before condition 1, the algorithm determines the i-th value of the increment of the momentum of $\alpha$-particle $dP_{\alpha_{i}}$:
\begin{equation}
    d\Vec{P}_{\alpha_{i}}=\Vec{F}^{e}_{\alpha_{i}}dt
    \label{eq}
\end{equation}
It corresponds to the termination of the program when the excess of $dT$ kinetic energy transferred to He exceeds the ionization potential of $O$~--helium $I_{0}$, which results in the destruction of the bound $O$~--helium system:
\begin{equation}
    dT<I_{0}\approx1.6 MeV
    \label{eq}
\end{equation}
\begin{equation}
    dT=\cfrac{dP_{\alpha_{i}}^2}{2m_{\alpha}}
    \label{eq}
\end{equation}
Condition 2 is described by formula (6) in the previous section.
As you can see from Figure 6, at each loop, the program calculates the total force acting on the $OHe$ system:
\begin{equation}
    \Vec{F}_{sum}=\Vec{F}^{e}_{ZO}+ \Vec{F}^{e}_{\alpha}
    \label{eq}
\end{equation}
With its help, the increment of the momentum $dP$ of $OHe$ system is calculated, which is, in the aggregate, the increment of the momentum of $O^{--}$.
\begin{equation}
    d\Vec{P}=\Vec{F}_{sum}dt
    \label{eq}
\end{equation}
Using the momentum increment $dP$, the $O^{--}$ velocity increment $dV$ is calculated for the subsequent finding of the new velocity used in the next iteration:
\begin{equation}
    d\Vec{V}=\cfrac{d\Vec{P}}{m_{O^{--}}+m_{\alpha}}
    \label{eq}
\end{equation}
The result of the algorithm is the reconstructed trajectories of $\alpha$-particle and $O^{--}$. One example is shown in Figure 6, where the blue circle shows the location of the target nucleus, the red asterisk and the purple square are the initial locations of the $\alpha$-particle and the $O^{--}$ particle, respectively, yellow dots and the green dashed line show the trajectories of the $\alpha$-particles and particles $O^{--}$ respectively. In the figure under consideration, one can observe the deviation of the trajectory $O^{--}$ from the initial direction, which is associated with the Coulomb interaction between the $He$ nucleus and the target nucleus. This happens because $He$ is closer to the origin and is repelled from the target nucleus more strongly than the $O^{--}$ particle is attracted to it.
\begin{figure}[h!]
\centering
\includegraphics[scale=0.35]{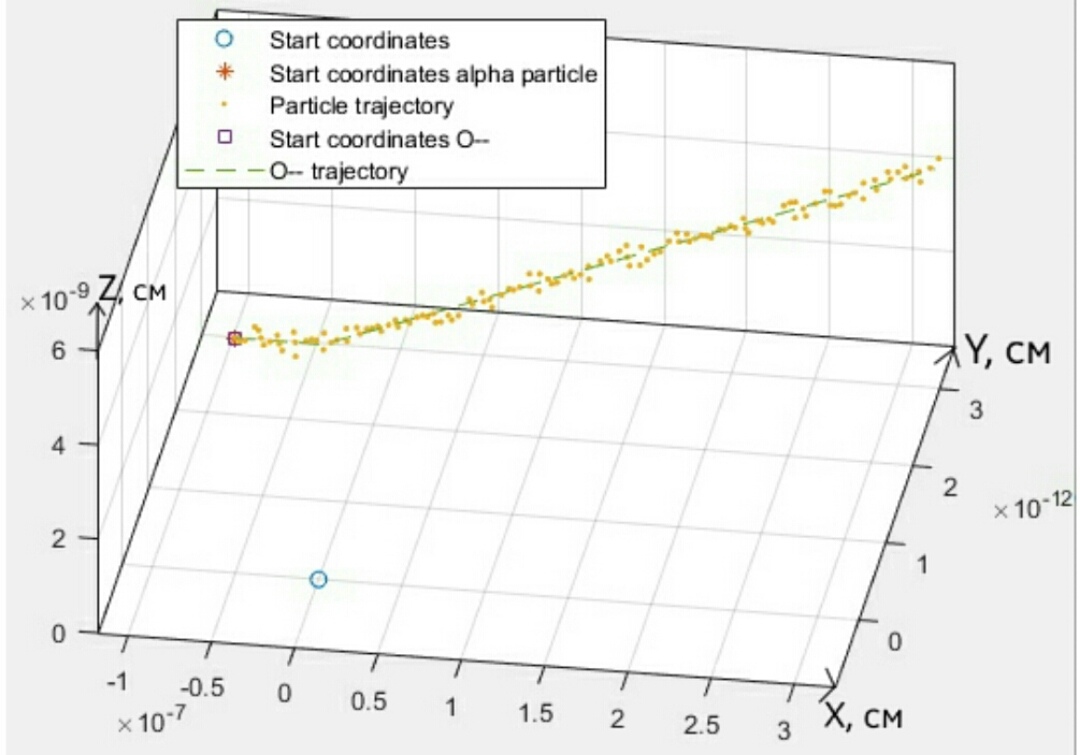}
\caption{$\alpha$-particle and particle $O^{--}$ trajectories}
\label{fig:lol1}
\end{figure}

\subsection{Nuclear interaction in the $OHe$~--nucleus system}
At this stage, the program was supplemented with a nuclear interaction of the Saxon-Woods type, between the $He$ nucleus and the target nucleus, determined by the force $\Vec{F}^{N}_{\alpha}$:
\begin{equation}
    \Vec{F}^{N}_{\alpha}=-\cfrac{\cfrac{U_{0}}{a}\exp{\biggl(\cfrac{r_{\alpha}-R_{Z}}{a}\biggr)}\cfrac{\vec{r}_{\alpha}}{r_{\alpha}}}{\Biggl(1+\exp{\biggl(\cfrac{r_{\alpha}-R_{Z}}{a}\biggr)}\Biggr)^2},
    \label{eq}
\end{equation}
where $R_{Z}$ is the radius of the target nucleus, $U_{0}$ is the depth of the potential well, $a$ is a constant parameter.

In this case, the total force acting on the system $OHe$, $\Vec{F}_{Sum}$, is now calculated as follows:
\begin{equation}
    \Vec{F}_{Sum}=\Vec{F}^{e}_{ZO}+\Vec{F}_{\alpha},
    \label{eq}
\end{equation}
where $\Vec{F}_{\alpha}$ is the total force acting on the $\alpha$-particle:
\begin{equation}
    \Vec{F}_{\alpha}=\Vec{F}^{e}_{\alpha}+\Vec{F}^{N}_{\alpha}
    \label{eq}
\end{equation}
Simulation is performed according to the algorithm described in the previous paragraph, where $dP_{\alpha}$, the increment of the $\alpha$-particle momentum, is now calculated as follows:
\begin{equation}
    d\Vec{P}_{\alpha}=\Vec{F}_{\alpha}dt
    \label{eq}
\end{equation}
Based on the data obtained, the program builds the trajectories of the $\alpha$-particle and the $O^{--}$. In Figure 7, which shows the result of the program, the blue circle shows the location of the target nucleus, yellow dots and the green dashed line show the trajectories of the $\alpha$-particle and the $O^{--}$ particle in the XY plane, respectively.
\begin{figure}[h!]
\centering
\includegraphics[scale=0.8]{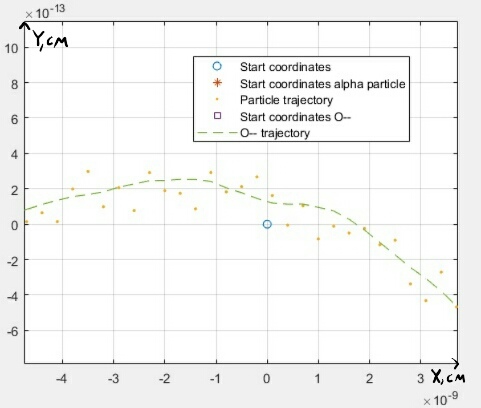}
\caption{The trajectory of $\alpha$-particle and the $O^{--}$ in the XY plane}
\label{fig:lol2}
\end{figure}

Figure 7 shows the effect of adding a nuclear force of interaction between the target nucleus and the $\alpha$-particle. Which consists in the fact that at small distances between particles, nuclear force can compensate for the effect of electromagnetic interaction. As a result, some beats are observed in the trajectory $O^{--}$.

%Figure 7 shows the difference in comparison with the trajectories of the $\alpha$-particle and $O^{--}$ shown in Figure 6. The difference appears due to the effect a nuclear force of interaction between the target nucleus and the $\alpha$-particle, which at small distances can compensate for the effect of electromagnetic interaction ~-- some beats are observed in the trajectory $O^{--}$.

%\subsection{Taking into account the final size of kernels}
%At this stage, a blank was made for future stages of work, in which it is supposed to take into account the finite sizes of nuclei by introducing the distribution of the density of nucleons (nuclear radius) and the density of protons (electromagnetic radius). For this, external functions were written in the program to extract this information based on empirical models for given (A, Z), that is, the number of nucleons A and the number of protons Z in the nucleus \cite{Seif:2015vca}.

%These external functions work as follows: when they are called, the values of the mass number A, the charge number Z, the density of protons or neutrons at the center of the nucleus $\rho_{o_{p,n}}$, the parameter of the quadrupole defect of the nucleus $\beta_{2}$, the values of which are taken from the database, and the distance from the center of the kernel r. Using the obtained data, the external function calculates and returns the value of the density of protons or neutrons at the point r ~-- $\rho_{p,n}$.

\section{Conclusions}
The advantage of the $OHe$ composite dark matter model is that it includes only one parameter of the "new" physics ~-- the $O^{--}$ mass. Atoms $OHe$ ~-- these neutral primary nuclear-interacting objects, provide the modern density of the dark matter and play the role of a non-trivial form of strongly interacting "dark" matter. Also, the $OHe$ hypothesis can explain the conflicting results of a direct search for "dark" matter, due to the specifics of the interaction of $O$~--helium with the substance of underground detectors. However, the correct quantum consideration of this model turns out to be rather difficult.

The $OHe$ hypothesis cannot work if no repulsive interaction occurs at some distance between $OHe$ and the nucleus, and the solution of this problem is vital for the further existence of the $OHe$ dark atom model \cite{Khlopov:2015nrq}.

Nuclear forces fall off exponentially, but they can be quite strong when the $OHe$ system comes close to the outer target nucleus. These are insignificant and insufficient distances for considering the $He$ nucleus as a point object.
%It is assumed that nuclear force can actually lead to a change in the polarization of $O$~--helium, which can lead to the creation of a dipole Coulomb barrier and that, as a result, a certain oscillatory system $ OHe $ ~-- nucleus should arise.
In this case, the perturbation theory can no longer be applied and it becomes rather problematic to solve the Schrödinger equation.
Therefore, the purpose of this work was to numerically simulate the interaction of the $OHe$ atom of "dark" matter with the nuclei of baryonic matter in order to reveal the conditions for the existence of their low-energy bound state and to calculate their effective interaction potential by a numerical method.

At the current stage, the our model describes a system of three point, interacting with each other through the Coulomb and nuclear forces, charged particles. The results of the work of the numerical model are the trajectories of motion of point particles entering the $OHe$ atom of dark matter, taking into account the electromagnetic and nuclear interactions between $O$~--helium and the target nucleus of baryonic matter, in the coordinate system $OHe$~--nucleus.

However, the process of numerical simulation has not yet been fully completed and in the future it is planned to improve it by introducing finite sizes of nuclei, by taking into account the distribution of the density of nucleons and the density of protons, and introducing the quantum-mechanical effect of tunneling $He$ nucleus into the nucleus of baryonic matter.
\section*{Acknowledgements}
The work by MK has been supported by the grant of the Russian Science Foundation (Project No-18-12-00213). The work by AM has been supported by the grant of the Russian Science Foundation, RSF 19-72-10161. 

%% The bibliography section

%section{\bibname}
%\printbibliography

\end{document}